\documentclass{pasj00}
\draft

\begin{document}
\SetRunningHead{Fujita et al.}{Hottest Cool Core Cluster}
\Received{2008/04/01}%{yyyy/mm/dd}
\Accepted{2008/06/23}%{yyyy/mm/dd}

%% ## `relaxed' is detelted.
\title{Suzaku Observation of the Ophiuchus Galaxy Cluster: One of the
Hottest Cool Core Clusters}

%%% begin:list of authors
\author{Yutaka \textsc{Fujita}\altaffilmark{1},
Kiyoshi \textsc{Hayashida}\altaffilmark{1},
Masaaki  \textsc{Nagai}\altaffilmark{1},
Susumu \textsc{Inoue}\altaffilmark{2}, \\
Hironori \textsc{Matsumoto}\altaffilmark{3},
Nobuhiro \textsc{Okabe}\altaffilmark{4},
Thomas.~H. \textsc{Reiprich}\altaffilmark{5}, \\
Craig~L. \textsc{Sarazin}\altaffilmark{6}, and	
Motokazu \textsc{Takizawa}\altaffilmark{7},
}
\altaffiltext{1}{Department of Earth and Space Science, Graduate School of
Science, \\ Osaka University, Toyonaka, Osaka 560-0043}
\email{fujita@vega.ess.sci.osaka-u.ac.jp}
\altaffiltext{2}{National Astronomical Observatory of Japan, 2-21-1 Osawa, Mitaka, Tokyo 181-8588}
\altaffiltext{3}{Department of Physics, Kyoto University, Kitashirakawa,
Sakyo-ku, Kyoto 606-8502}
\altaffiltext{4}{Astronomical Institute, Graduate School of Science,
Tohoku University, Sendai 980-8578}
\altaffiltext{5}{Argelander Institute for Astronomy (AIfA), Bonn
University, Auf dem H\"ugel 71, 53121 Bonn, Germany}
\altaffiltext{6}{Department of Astronomy, University of Virginia,
P.O. Box 400325, Charlottesville, VA 22904-4325, USA}
\altaffiltext{7}{Department of Physics, Yamagata University, Yamagata
990-8560}
%%% end:list of authors

%% `\KeyWords{}' always has to be placed before `\maketitle'.
 %Do NOT move this preamble from here!

\maketitle

\begin{abstract}
 We present the analysis of a Suzaku observation of the Ophiuchus galaxy
 cluster. We confirmed that the cluster has a cool core. While the
 temperature of the intracluster medium (ICM) decreases toward the
 center, the metal abundance increases. Except for the core ($r\lesssim
 50$~kpc), the cluster is hot ($\sim 9$--10~keV) and is almost
 isothermal for $r\lesssim 1$~Mpc; the latter contradicts a previous
 study. We do not detect the variation of the redshift of the ICM in the
 cluster; the upper limit of the velocity difference is $3000\:\rm km\:
 s^{-1}$.  The iron line ratios in X-ray spectra indicate that the ICM
 has reached the ionization equilibrium state. From these results, we
 conclude that the Ophiuchus cluster is not a major merger cluster but
 one of the hottest clusters with a cool core. We obtain the upper limit
 of non-thermal emission from the cluster, which is consistent with both
 the recent claimed detection with INTEGRAL and the recent upper limits
 with the Swift/BAT. If the cluster has bright non-thermal emission as
 suggested by the INTEGRAL measurement, it is probably not due to a
 recent major cluster merger.
\end{abstract}

\section{Introduction}

The cold dark matter (CDM) model predicts that the typical mass of
virialized objects in the Universe increases with time. This means that
the most massive or hottest galaxy clusters ($kT\gtrsim 10$~keV) should
be the objects that have formed most recently. In fact, X-ray
observations have shown that the hottest clusters are often
`major-merger' clusters. These clusters are now growing through the
mergers. For example, cluster 1E~0657--56 ($kT=14.8_{-1.2}^{+1.7}$~keV)
has a clear bow shock associated with a merger; a smaller cluster is
penetrating a larger cluster with a velocity of 3000--$4000\rm\: km\:
s^{-1}$ \citep{mar02}. On the other hand, these hottest clusters are
rarely `cool core clusters', which have a distinct cool core and are
almost isothermal outside the core. This is probably because cool cores
are destroyed by cluster mergers \citep{ric01,fuj02}.

Cluster mergers may accelerate relativistic particles at shocks and in
turbulent regions in the ICM
\citep{jaf77,rol81,sch87,sar99,tak00,fuj01,bru01,min01,ohn02,fuj03,ryu03,bru05,cas05,ino05,kan07}. With
BeppoSAX, non-thermal hard X-ray emission from relativistic particles
has been detected from several clusters such as the Coma cluster
\citep{fus04}. Most of them are actually merging clusters
\citep{nev04}. However, the detections are still controversial
\citep{ros04}, and confirmation with other instruments would be highly
desirable.  The Suzaku HXD has an improved sensitivity for hard X-rays
\citep{tak07}, and attempts have been made to detect non-thermal hard
X-ray emission from clusters. So far, no firm detections have been
reported \citep{kaw08}.

The Ophiuchus cluster is one of the hottest known clusters. It has been
observed with several satellites, such as HEAO~1 \citep{joh81}, EXOSAT
\citep{arn87}, Tenma \citep{oku88}, and GINGA \citep{kaf92}. Previous
observations with ASCA and BeppoSAX showed that the temperature of the
ICM is $\sim 10$~keV \citep{wat01,nev04}. In addition to the high
temperature, this cluster has a few interesting features. First, it has
been reported to have a complicated temperature distribution, in
particular, the temperature of its western part reaches $\sim 20$~keV
\citep{wat01}. This temperature is comparable to that observed in
1E~0657--56 \citep{mar02}. Thus, this cluster may be undergoing a major
merger that induces the motion of the ICM with a velocity of $\gtrsim
3000$ km s$^{-1}$.  Second, non-thermal hard X-ray emission may have
been detected from the cluster.  \citet{nev04} searched for non-thermal
emission with BeppoSAX and found some evidence at a $2\sigma$
level. Recently, \citet{eck08} detected non-thermal emission from the
center of the cluster with INTEGRAL at a $6.4\sigma$ level.  However, a
recent long exposure with the Swift/BAT instrument failed to detect any
nonthermal emission, and set an upper limit below the claimed INTEGRAL
detection \citep{aje08,oka08}.

In this paper, we report our observational results of the Ophiuchus
cluster with Suzaku. Thanks to the low background of Suzaku
\citep{mit07}, we can obtain an X-ray spectrum even in regions far from
the cluster center; we discuss the temperature and metal abundance
distributions in the ICM. The high spectral resolution of the XIS
\citep{koy07} on board Suzaku also enables us to measure or limit the
bulk motion of the ICM. Moreover, the hard X-ray detector, HXD, can
constrain the non-thermal emission from the cluster, which can be
compared with the INTEGRAL and Swift/BAT results.

In this paper, we assume cosmological parameters of $\Omega_0=0.3$,
$\lambda_0=0.7$, and $H_0=70\rm\: km\: s^{-1}\: Mpc^{-1}$. Since the
cluster redshift is $z=0.028$ \citep{lah89}, one arc-minute corresponds
to 33~kpc. All statistical errors are given at the 90\% confidence level
unless otherwise mentioned.

\section{Observations}

\begin{figure}
  \begin{center}
    \FigureFile(100mm,100mm){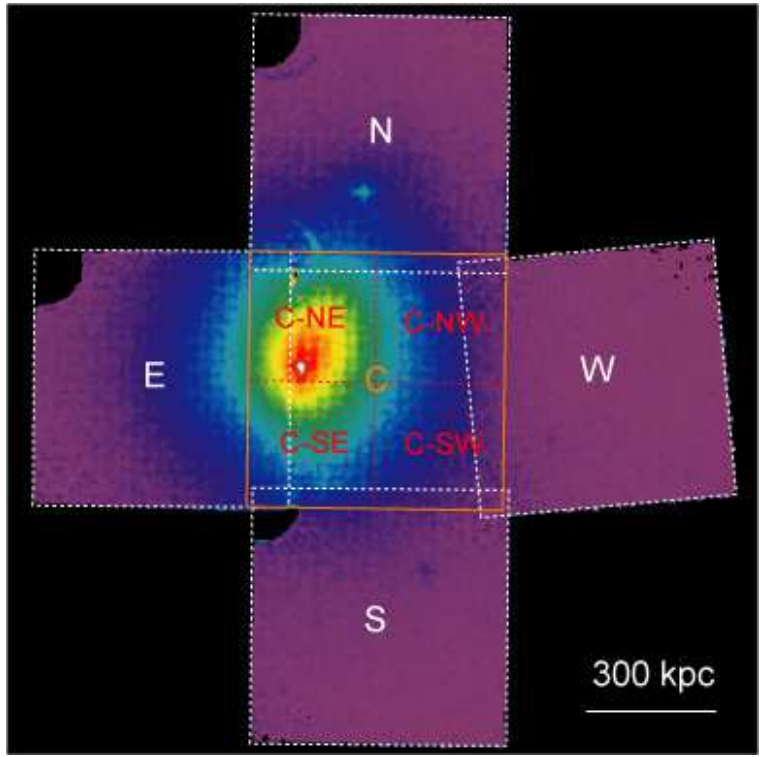}
  \end{center}
  \caption{A mosaic image of the Ophiuchus cluster obtained with Suzaku
 XIS. For each of the five pointings, the images from the three XIS
 detectors (XIS0, XIS1, and XIS3) were combined.  The image is corrected
 for vignetting but not for background.  The calibration sources at the
 corners of the fields were excluded. Regions used for spectral analysis
 are shown.}\label{fig:image}
\end{figure}

We observed the Ophiuchus cluster with Suzaku on 2007 March 21--24 for a
total exposure time of 105~ks. Figure~\ref{fig:image} shows a mosaic
image of the cluster obtained with the XIS; the field of view (FOV) of
the XIS is \timeform{17'.8}$\times$\timeform{17'.8}. We observed five
regions with the XIS: C (central), N (north), W (west), S (south), and E
(east) regions. The exposure time for region~C is 15~ks, and those for
other regions are 22.5~ks. The XIS was operated in the normal full-frame
clocking mode. The edit mode was $3\times 3$ and $5\times 5$ and we
combined the data of both modes for our analysis. We employed the
following calibration files: HXD (20071201), XIS (20071122), and XRT
(20070622). Events with ASCA grades of 0, 2, 3, 4, and 6 were
retained. We excluded the data obtained at the South Atlantic Anomaly,
during Earth occultation, and at the low elevation angles from Earth rim
of $<5\degree$. For the XIS data, we further excluded the data with
elevation angle from the rim of the shining Earth smaller than
$20\degree$. For the HXD data, we removed the data obtained at the
locations where the Cut-Off-Rigidity (COR) is lower than 6~GV.

\section{Spectral Analysis}
\label{sec:spec}

Our spectral analysis is similar to that in \citet{fuj08}. We extract
X-ray spectra for the five regions (C, W, S, E, and N) shown in
figure~\ref{fig:image}. For the central region (C), we also analyze the
square subsections (C-NW, C-SW, C-SE, and C-NE). We calculated the
effective area for each XIS chip using {\sc xissimarfgen}, which
provides the ancillary response file (ARF) through Monte Carlo
simulations \citep{ish07}. All spectra were rebinned to give a minimum
of 50 raw counts per spectral bin to allow $\chi^2$ statistics to be
applied.  The non-X-ray instrumental background (NXB) spectrum is
subtracted from the spectrum of each region. The NXBs for individual
regions were constructed from night earth data and were generated by the
routine {\sc xisnxbgen} included in Suzaku FTOOLS Ver.\ 7, including the
time-variation of the NXB \citep{taw08}.  We always include the NXB in
all of the following spectral fits.

\begin{figure}
  \begin{center}
    \FigureFile(100mm,100mm){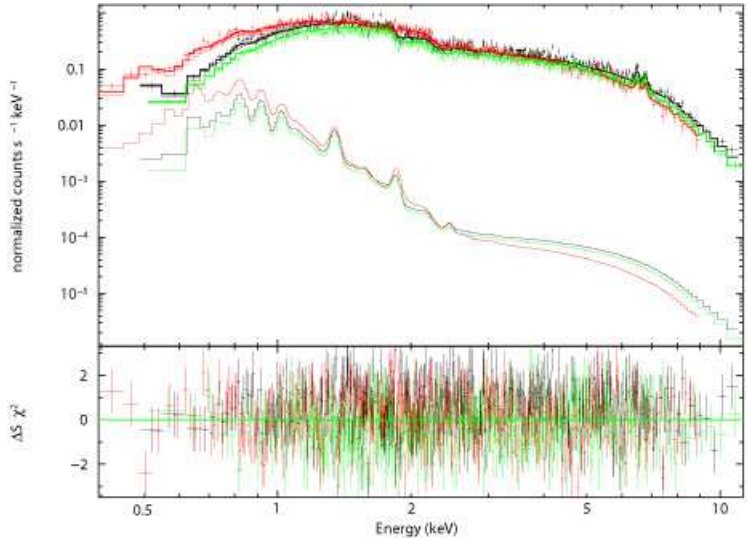}
  \end{center}
  \caption{The XIS spectra for region C-NW (crosses). The results of the
fit are shown by the solid lines; the lower lines are the background
emission (CXB plus the Galactic one). The lower panel plots the
residuals divided by the $1\sigma$ errors. The NXB spectra have been
subtracted.}\label{fig:spec}
\end{figure}

Using XSPEC (Ver.\ 12), the spectra of the three XISs are simultaneously
fitted with a single thermal model (APEC) representing the ICM and with
the Galactic absorption (WABS; $N_{\rm H}$). The parameters of these
models are set to be the same for all the XISs except for the
normalization of the APEC model, which is the main component of the
spectra. While the normalizations for the two front-illuminated (FI)
chips (XIS~0 and 3) are the same, that for the back-illuminated (BI)
chip (XIS~1) is allowed to have an independent value, because of the
residual calibration inaccuracies between FI and BI. In the fits, we
also consider the contributions of the cosmic X-ray background (CXB) and
the Galactic soft X-ray emission. The CXB spectrum is given by a
power-law with an index of 1.412 and the flux in the 2--10~keV band is
$F_{\rm CXB}=6.38\times 10^{-8}\rm\: erg\: cm^{-2}\: s^{-1}\: sr^{-1}$
\citep{kus02}. We call this CXB spectrum `the normal CXB spectrum'.  The
Galactic soft X-ray emission model is constructed on the model of
\citet{sno98}\footnote{http://heasarc.gsfc.nasa.gov/cgi-bin/Tools/xraybg/xraybg.pl}.
The emission consists of two thermal components with temperatures of $kT
\approx 0.1$ keV (the local hot bubble) and 0.3~keV (the Milky Way
halo); the values are slightly dependent on the region observed. The
metal abundance and the redshift of both components are solar and zero,
respectively. The parameters for the Galactic soft X-ray emission are
fixed in the fits except for the overall normalization; the relative
normalization between the two components is fixed at cooler/warmer
$\approx 0.15$, which is provided by the Galactic soft X-ray emission
model$^1$.  While the ARFs for the ICM emission were constructed
assuming that it follows a beta model for the cluster ($\beta=0.747$ and
the core radius of $6'$; \cite{rei02}), those for the background
emission (CXB plus the Galactic one) were made assuming that the
emission is spatially uniform. Figure~\ref{fig:spec} shows the spectra
of the three XISs for a region with typical photon counts for the
spectral analysis (C-NW) and the result of the fit.

\begin{figure}
  \begin{center}
    \FigureFile(80mm,80mm){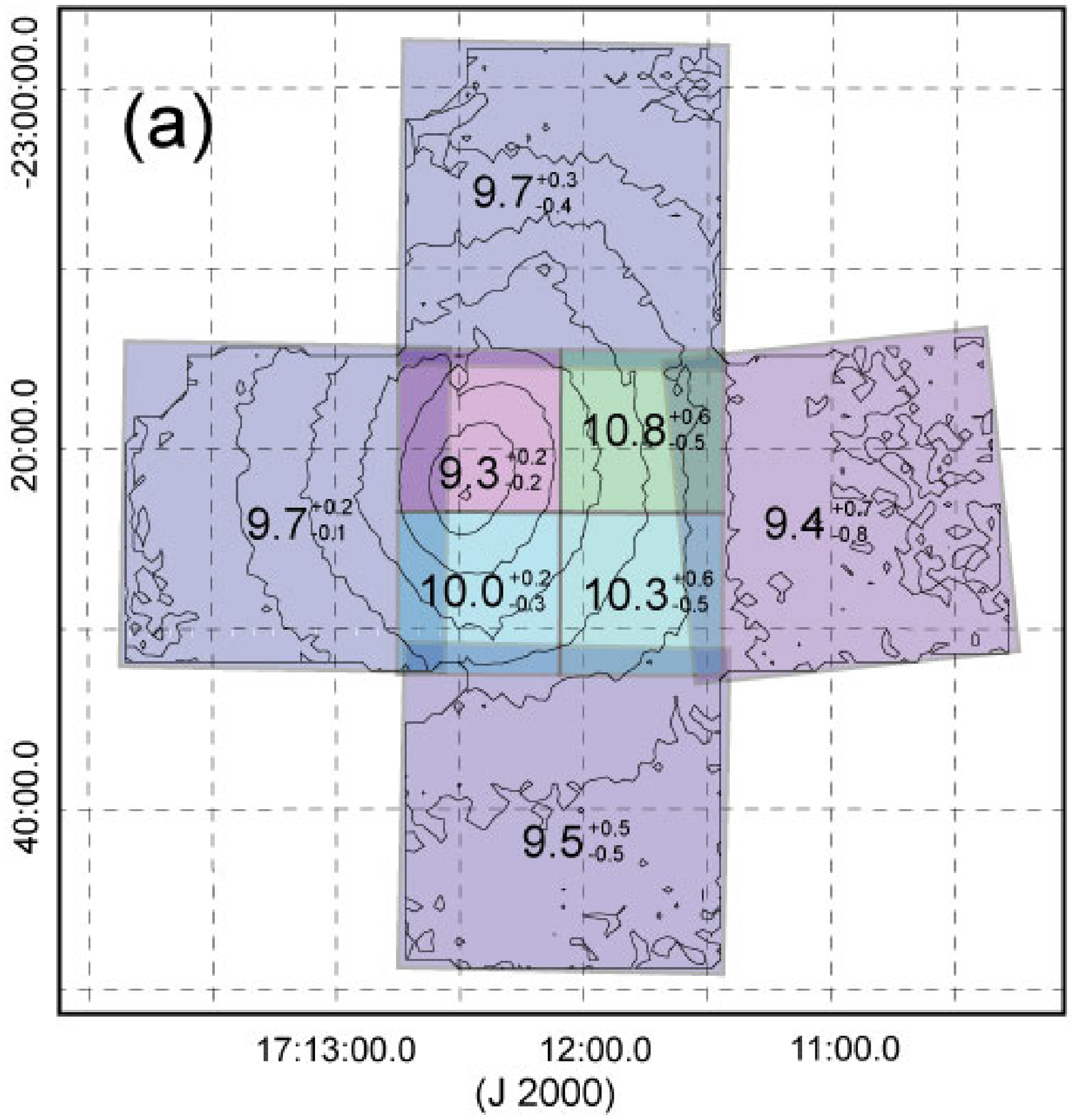}\FigureFile(80mm,80mm){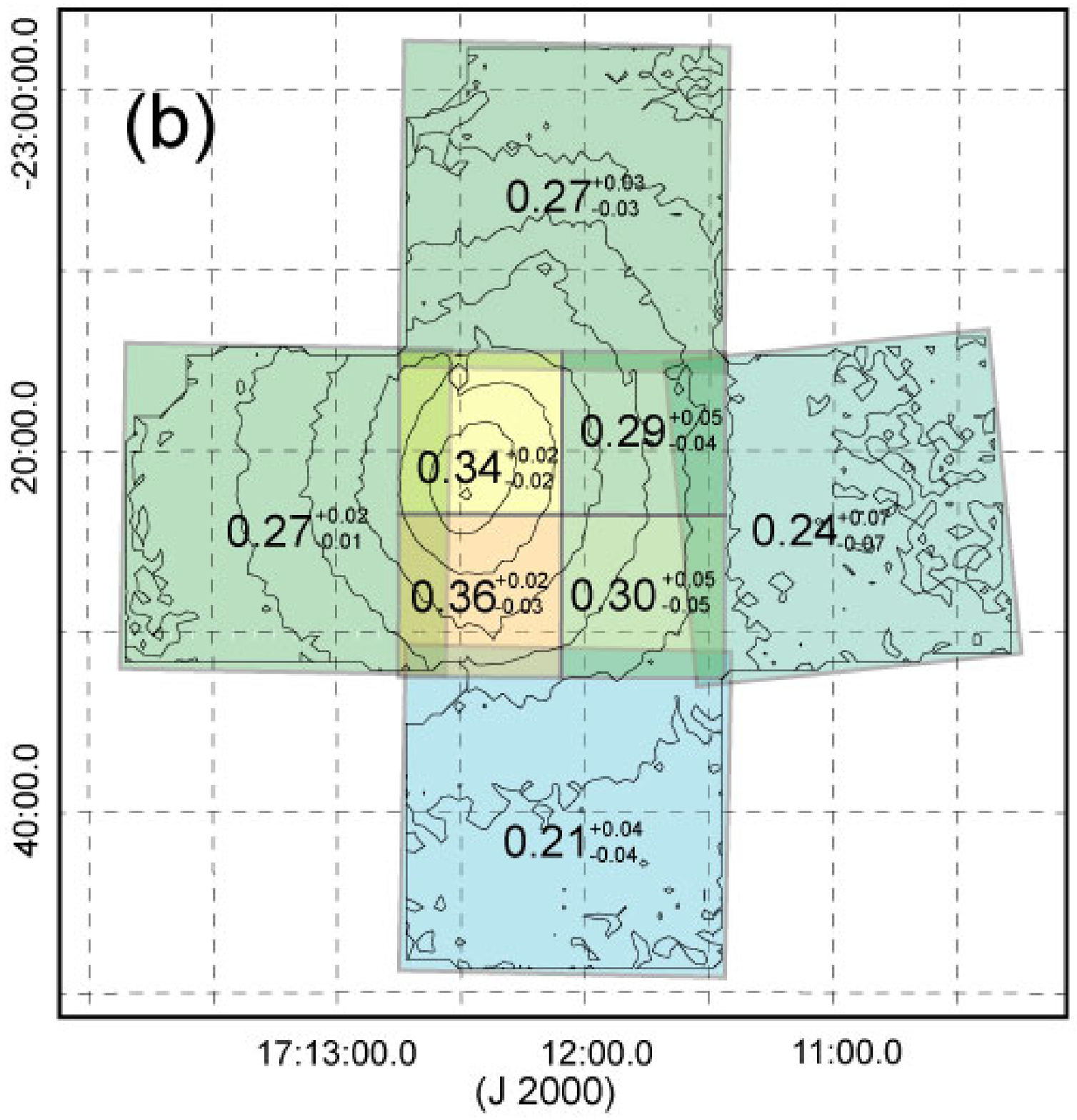}
  \end{center}
  \caption{(a) Temperature map of the Ophiuchus cluster. The numbers
 represent the temperatures (keV) for each region. X-ray brightness
 contours (logarithmically spaced by a factor of 2) are overlaid. (b)
 Same as (a) but for metal abundance ($\rm Z_\odot$)}\label{fig:TZ}
\end{figure}

Table~\ref{tab:spec} shows the results of the spectral
fits. Temperatures ($T$) and abundances ($Z$) are also shown in
figure~\ref{fig:TZ}.  In determining $T$, $Z$, and the absorbing column
$N_{\rm H}$, the redshift of the ICM is fixed at the optical value of
$z=0.028$ \citep{lah89}.  The spectral fits show that the cluster is
almost isothermal with a temperature of $kT\sim 9$--10~keV and an
abundance of $Z\sim 0.3\:\rm Z_\odot$. When we derive the errors, we
vary the normalization of the CXB in order to include possible
field-to-field variance. \citet{kus02} derived a $1\sigma$ deviation of
the CXB intensity to be $(6.49^{+0.56} _{-0.61} )$\% using ASCA GIS
data. We roughly estimate that the CXB uncertainty for the XIS field is
$6.49\sqrt{0.4/0.088} \sim 14$\%, considering the typical size of the
area studied in \citet{kus02} ($\sim 0.4\rm\: deg^2$) and the FOV of the
XIS ($\sim 0.088\rm\: deg^2$). Note that this value depends on the flux
threshold to resolve point sources. While \citet{kus02} employed
$2\times 10^{-13}\rm\: erg\: s^{-1}\: cm^{-2}$ (2--10~keV), we have
excluded point sources with smaller fluxes. Thus, the actual uncertainty
might be smaller. We fit the XIS spectra of regions C, N, W, S and E for
14\% higher or lower CXB normalizations than that of the normal CXB
spectrum and derive temperatures and metallicities. Adding linearly the
statistical errors of the fits to this CXB uncertainty, we obtain the
errors. Similarly, we assume $\sim 28$\% uncertainty for the CXB
normalizations for the four smaller regions in C. In
table~\ref{tab:spec}, $N_{\rm H}$ shows a large variation.  Given that
such large variations and values are not seen in clusters at higher
Galactic latitudes, this variation is likely to be Galactic.

For comparison, we also present the results of previous studies in
table~\ref{tab:spec}.  The ASCA data are for the central region of the
cluster (region~2 in \cite{wat01}). They are consistent with the Suzaku
results for region~C.  The HXD results will be discussed in
section~\ref{sec:nonth}.

\begin{table}
  \caption{The results of spectral fits and redshifts.}\label{tab:spec}
  \begin{center}
    \begin{tabular}{cccccc}
  \hline              
    & $kT$ & $Z$ & $N_{\rm H}$ & $\chi^2/$dof & $z$\\ 
  Region  & (keV)  & ($\rm Z_\odot$) & 
($10^{21}\:\rm cm^{-2}$) & 
 & \\
  \hline
  C & $9.7_{-0.2}^{+0.1}$ & $0.33_{-0.01}^{+0.02}$ & $3.1_{-0.0}^{+0.1}$ & 5399.08/5044 & ..... \\
  C-NW & $10.8_{-0.5}^{+0.6}$ & $0.29_{-0.04}^{+0.05}$ & $2.5_{-0.1}^{+0.2}$ & 1326.70/1176 & $0.0319\pm 0.0020$\\ %RD
  C-SW & $10.3_{-0.5}^{+0.6}$ & $0.30_{-0.05}^{+0.05}$ & $2.8_{-0.3}^{+0.2}$ & 1042.62/936 & $0.0292\pm 0.0028$\\ %LD
  C-SE & $10.0_{-0.3}^{+0.2}$ & $0.36_{-0.03}^{+0.02}$ & $3.2_{-0.1}^{+0.1}$ & 2935.76/2602 & $0.0290\pm 0.0013$ \\ %LU
  C-NE & $9.3_{-0.2}^{+0.2}$ & $0.34_{-0.02}^{+0.02}$ & $3.3_{-0.0}^{+0.1}$ & 3768.32/3732 & $0.0297\pm 0.0011$	\\ %RU
  N    & $9.7_{-0.4}^{+0.3}$ & $0.27_{-0.03}^{+0.03}$ & $2.4_{-0.1}^{+0.1}$ & 2425.38/2221 & $0.0309\pm 0.0017$	\\
  W    & $9.4_{-0.8}^{+0.7}$ & $0.24_{-0.07}^{+0.07}$ & $1.4_{-0.2}^{+0.2}$ & 909.52/877 & $0.0287\pm 0.0054$	\\
  S    & $9.5_{-0.5}^{+0.5}$ & $0.21_{-0.04}^{+0.04}$ & $1.8_{-0.1}^{+0.2}$ & 1499.63/1375 & $0.0283\pm 0.0025$	\\
  E    & $9.7_{-0.1}^{+0.2}$ & $0.27_{-0.01}^{+0.02}$ & $2.9_{-0.0}^{+0.1}$ & 4187.73/4028 & $0.0305\pm 0.0011$	\\
  ASCA & $10.0_{-0.3}^{+0.3}$ & $0.34_{-0.04}^{+0.04}$ & $3.0$ (Fixed) & 66.6/45 & ..... \\
  BeppoSAX & $9.1_{-0.5}^{+0.6}$ & $0.49_{-0.08}^{+0.08}$ & ..... & ..... & ..... \\
  Suzaku HXD & $9.0_{-0.3}^{+0.3}$ & $0.3$ (Fixed) & 0.0 (Fixed) & 133.92/152 & ..... \\
  \hline
    \end{tabular}
  \end{center}
\end{table}

\section{Deprojection Analysis}
\label{sec:depro}

The Chandra image of the Ophiuchus cluster exhibits cold fronts around
the core \citep{asc06}. Numerical simulations have shown that the region
surrounded by cold fronts has a low temperature compared with the
outside of the cold fronts \citep{fuj04,asc06}, which is consistent with
observations for several clusters (e.g. \cite{mar01,maz08}). In order to
study the temperature and metal abundance of the core in detail, we made
a deprojection analysis to remove projection effects. The deprojection
procedure we adopted is the same as that in \citet{bla03} and
\citet{fuj04f}.

\begin{figure}
  \begin{center}
    \FigureFile(100mm,100mm){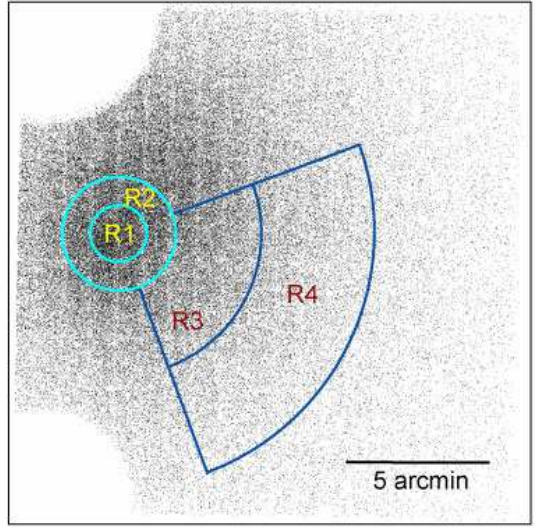}
  \end{center}
  \caption{An XIS~0 sky image of region~C. The regions used for a
 deprojection analysis are shown.}\label{fig:ann}
\end{figure}

We choose annuli and sectors centered on the X-ray peak as shown in
figure~\ref{fig:ann}. We fix the Galactic absorption at the value of
region C ($N_{\rm H}=3.1\times 10^{21}\rm\: cm^{-2}$;
table~\ref{tab:spec}). The Galactic soft X-ray emission is also fixed at
the values of region C. The CXB is fixed at the normal value. These
background models are always included in the following procedure.

First, we fit the spectrum from region R4 with a single ICM model
(APEC). Then, the next annulus (R3) is fitted. The model used for this
annulus is a combination of the best-fitting model of the exterior
annulus with the normalization scaled to account for the spherical
projection of the exterior shell onto the inner one, along with another
APEC component added to account for the emission at the radius of
interest. This process is continued inward, and we fitted one spectrum
at a time.  The FI and BI chip are treated separately because of their
different spectral responses.

\begin{figure}
  \begin{center}
    \FigureFile(80mm,80mm){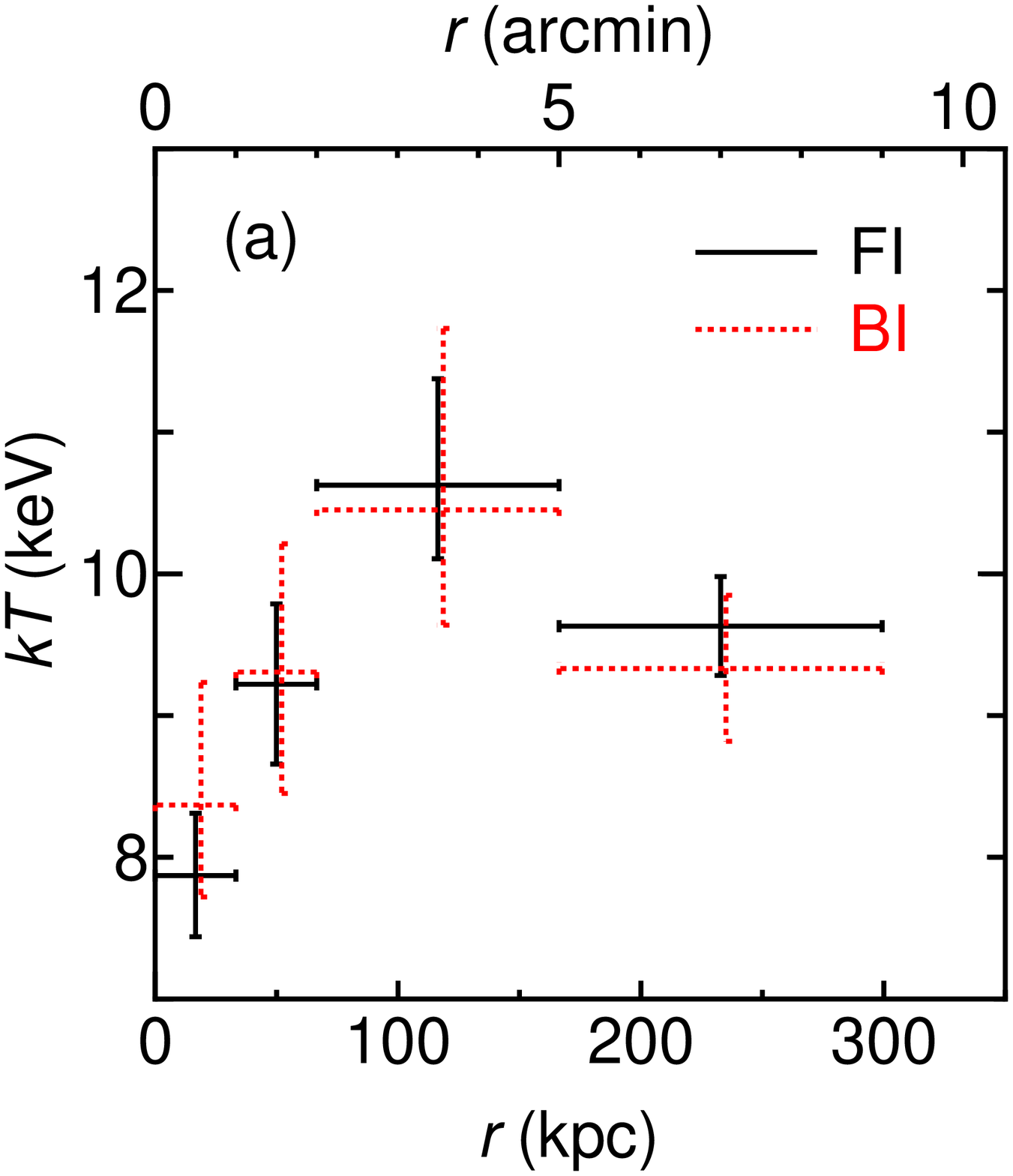}\FigureFile(80mm,80mm){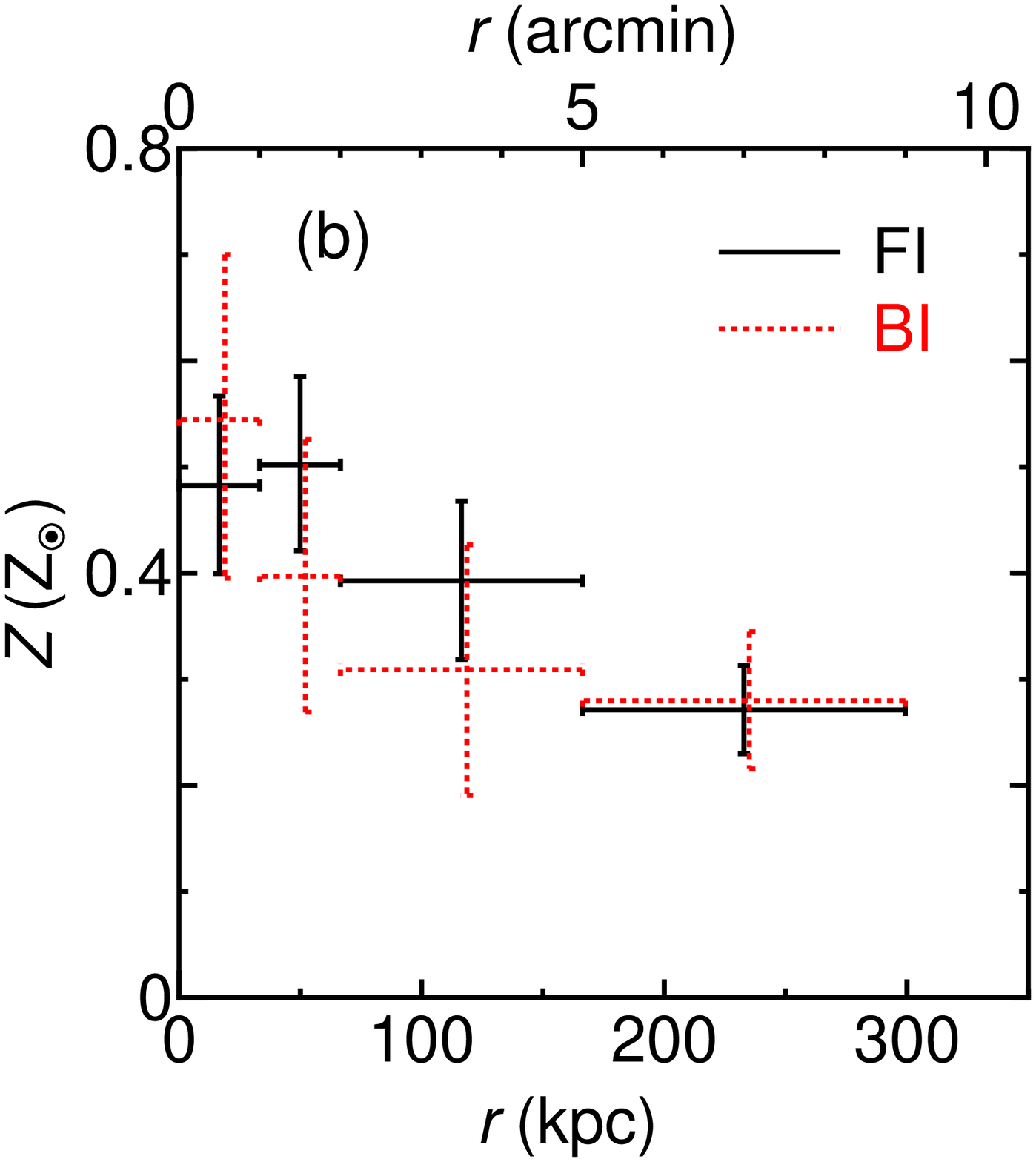}
  \end{center}
  \caption{(a) Temperature and (b) metal abundance profiles as functions
 of the distance from the cluster center.}\label{fig:depro}
\end{figure}

The results of the fits are shown in figure~\ref{fig:depro}. The
temperature of the ICM decreases to $kT\sim 8$~keV toward the cluster
center, while the abundance increases to $Z\sim 0.5$--$0.6\rm\:
Z_\odot$. This behavior is often found in cool core clusters
\citep{fuk94,fin01,deg01,poi05,vik05}. These results are conservative
because, considering point-spread-function (PSF) smearing by the angular
resolution of Suzaku ($\sim 2'$ half power diameter, independent of
energy), the actual central temperature and abundance will be even lower
and higher, respectively.

\section{Bulk Motion of the ICM}
\label{sec:bulk}

By measuring the redshifts of metal lines in X-ray spectra, we can
estimate the velocity of the ICM.  There have been a number of previous
attempts to measure bulk motions in the ICM using X-ray line shifts
(e.g., \cite{dup06}).  Since the XISs on board Suzaku have an excellent
spectral resolution, they are suitable for such studies \citep{ota07}.
In this section, we estimate the bulk motion of the ICM of the Ophiuchus
cluster.

\begin{figure}
  \begin{center}
    \FigureFile(80mm,80mm){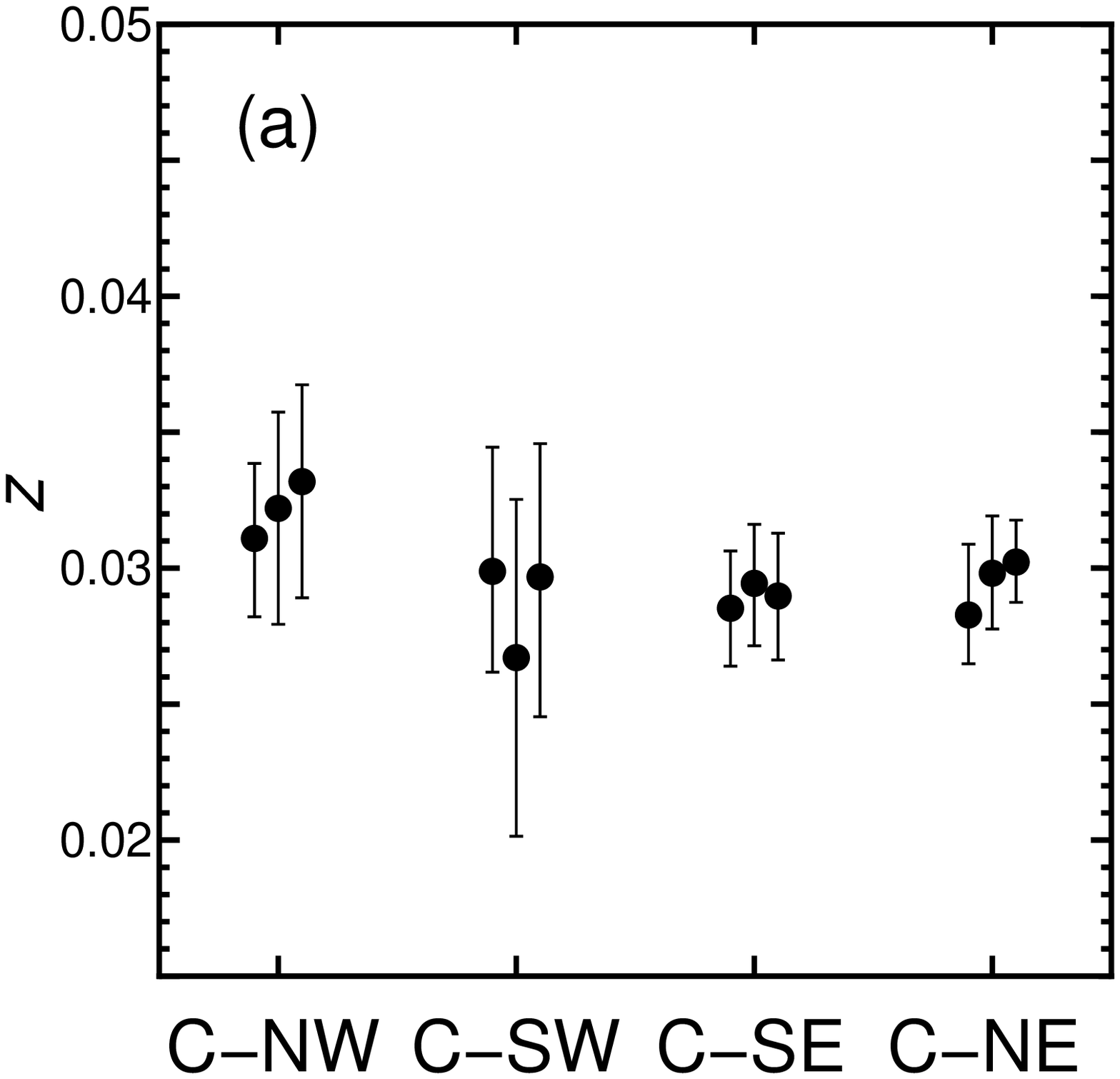}\FigureFile(80mm,80mm){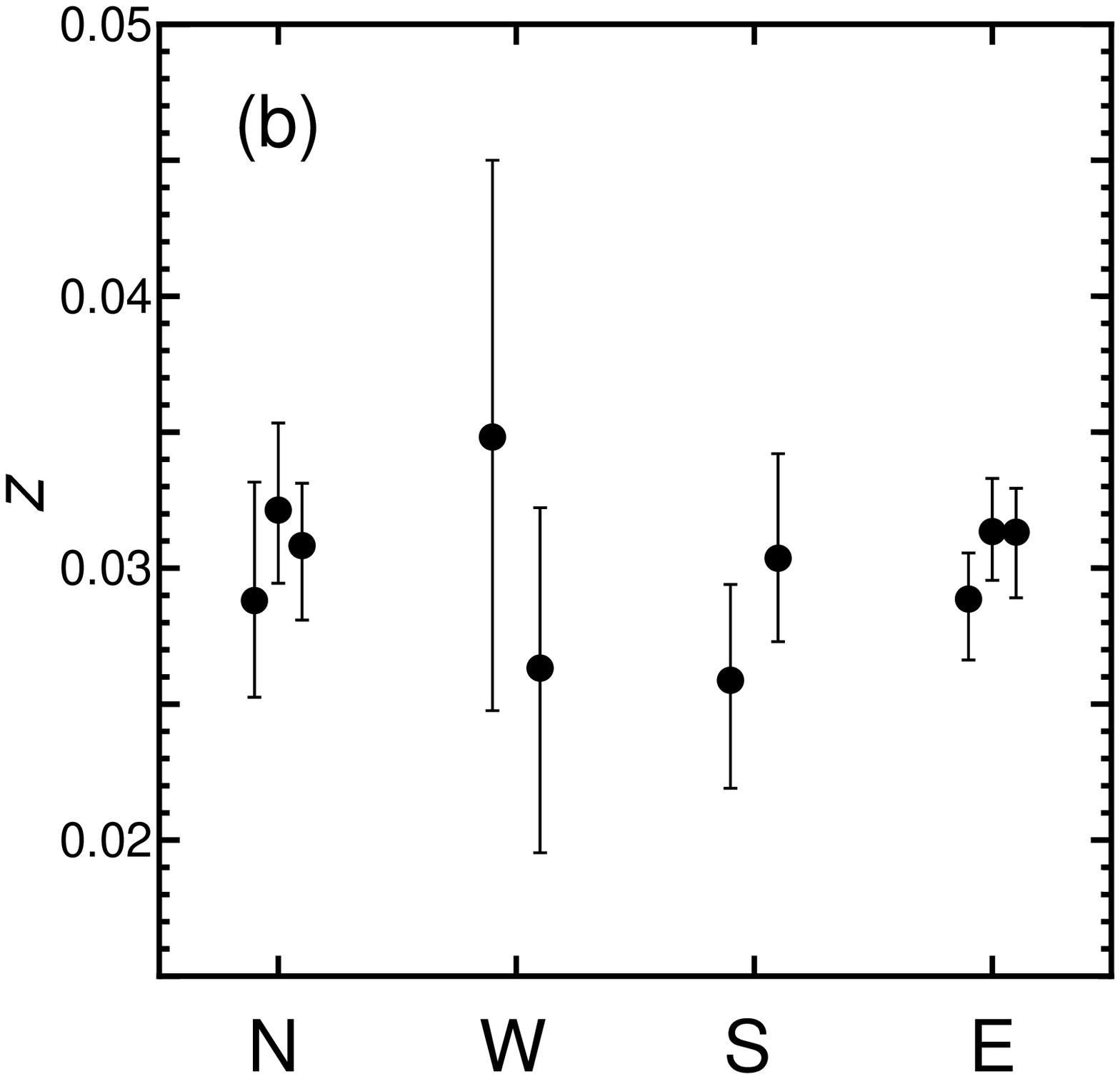}
  \end{center}
  \caption{Redshifts of the ICM in the (a) central regions and (b)
 surrounding regions. The three filled circles for each region are the
 gain corrected redshifts ($z_{\rm gain}$) for XIS~0, 1, and 3 from left
 to right. For regions W and S, only data from the FI chips (XIS~0, and
 3) are shown.  }\label{fig:redshift}
\end{figure}

We investigate the redshifts of the ICM for four central regions (C-NW,
C-SW, C-SE, and C-NE) and the four surrounding regions (N, W, S, and E)
shown in figure~\ref{fig:image}. A `gain correction' of the CCD chips in
instruments like the XIS must be done carefully for a precision
measurement of the redshift of the metal lines.  The XISs are always
illuminated by calibration sources ($^{55}\rm Fe$) that appear as a Mn
K$\alpha$ line (5.895~keV) in spectra if not masked. In this section, we
treat the three XIS chips separately in order to allow for the gain
difference among them.  First, we limited the energy range to 5--8~keV
and fit the ICM spectra including the emission from the calibration
sources with bremsstrahlung and gaussian line components.
The Mn K$\alpha$ line from the calibration sources and the
Fe~K lines from the ICM are each fitted with the gaussian
components. 
We put the gaussian components for the Fe K lines just because they may
influence the line center of Mn K$\alpha$ line; we do not use them for
the determination of redshift. The central energy of the Mn K$\alpha$
line determined by this fit is referred as $E_{\rm obs}(\rm Mn\;
K\alpha)$. The gain correction factor is given by
\begin{equation}
 f_{\rm gain}=\frac{E_{\rm
obs}(\rm Mn\; K\alpha)}{E_0(\rm Mn\; K\alpha)}\:,
\end{equation}
where $E_0(\rm Mn\; K\alpha)$ is the true energy (5.895~keV). Typically,
$|f_{\rm gain}-1|\lesssim 0.001$, 
and the uncertainty of $f_{\rm gain}$ is $\sim 0.001$. On the other
hand, we determine the redshift of the ICM, $z_{\rm fit}$, through a
spectral fit that is the same as that in section~\ref{sec:spec} but for
each chip. Thus, a gain-corrected redshift is given by
\begin{equation}
 z_{\rm gain}=f_{\rm gain}(z_{\rm fit}+1)-1 \:.
\end{equation}
We present $z_{\rm gain}$ for each region in
figure~\ref{fig:redshift}. The values averaged for the three XISs are
shown in table~\ref{tab:spec}. For regions W and S, only data for FI
chips (XIS~0, and 3) are shown because the BI chip (XIS~1) is less
sensitive to hot gas ($\gtrsim 2$~keV) and the redshifts cannot be
constrained. The redshifts in figure~\ref{fig:redshift} are consistent
with being constant and the average is $z=0.030\pm 0.003$, which agrees
with the optical redshift of the cluster ($z=0.028$). The difference of
the maximum (C-NW) and minimum redshift (W) in table~\ref{tab:spec} is
$\lesssim 0.01$, which means that the velocity difference between
different regions is $\lesssim 3000\rm\: km\: s^{-1}$.

\section{Non-Thermal Hard X-ray Emission}
\label{sec:nonth}

The Suzaku HXD PIN detector is sensitive to hard X-rays ($\gtrsim
10$~keV) and has a wider FOV ($34'\times 34'$ full width at half
maximum, FWHM) than that of the XIS
(\timeform{17'.8}$\times$\timeform{17'.8}). Although the center of the
HXD field is shifted from that of the XIS field by $\sim
\timeform{3'.5}$, the shift is much smaller than the size of the HXD
field.  Thus, all of the five HXD fields (corresponding to the XIS
fields C, N, W, S, and E) cover at least partially the central region of
the cluster, where INTEGRAL detected hard X-ray excess emission
\citep{eck08}.

After correcting for dead time, the exposure times of the five fields
are 12.6, 18.1, 19.4, 20.0, 20.1~ks for C, N, W, S, and E, respectively.
We fit the HXD spectra with a thermal model (APEC), which represents the
ICM, as well as a CXB component. We first simultaneously fit the spectra
of the five fields. A redistribution matrix file for the HXD nominal
position is used for the ICM emission. The ARFs of the ICM emission for
each spectrum are generated with {\sc hxdarfgen} and {\sc addarf} by
assuming that the emission follows the same beta model that we adopted
for the XIS (section~\ref{sec:spec}). We fix the metal abundance of the
APEC model at $0.3\rm\: Z_\odot$ and the redshift at $z=0.028$ (optical
redshift). The CXB is assumed to be uniform in a $2\degree\times
2\degree$ field, and a corresponding redistribution matrix file is
employed. The spectrum of the CXB is represented by a power-law model
multiplied by a HIGHECUT model in XSPEC. We adopt the power-law index of
1.29, the power-law normalization of $9.412\times 10^{-9}\rm\: photons\:
cm^{-2}\: s^{-1}\: FOV^{-1}\: keV^{-1}$, the cutoff energy of $1\times
10^{-4}$~keV, and the e-folding energy of 40~keV, following a standard
model\footnote{http://heasarc.nasa.gov/docs/suzaku/analysis/pin\_cxb.html}. Thus,
the free parameters for the fit are the temperature and normalization of
the APEC model. While the temperature is common, the normalization is
allowed to vary among the five spectra. We do not consider the Galactic
absorption because it does not affect the results in the HXD energy
band. The NXB spectra created from the pinnxb\_ver2.0\_tuned NXB model
are employed as a normal NXB. We find that the five spectra can be
fitted with the thermal model with a temperature of
$kT=8.9_{-0.4}^{+0.2}$~keV. The normalizations of the five spectra agree
with each other within $\pm 10$\%.

\begin{figure}
  \begin{center}
    \FigureFile(100mm,100mm){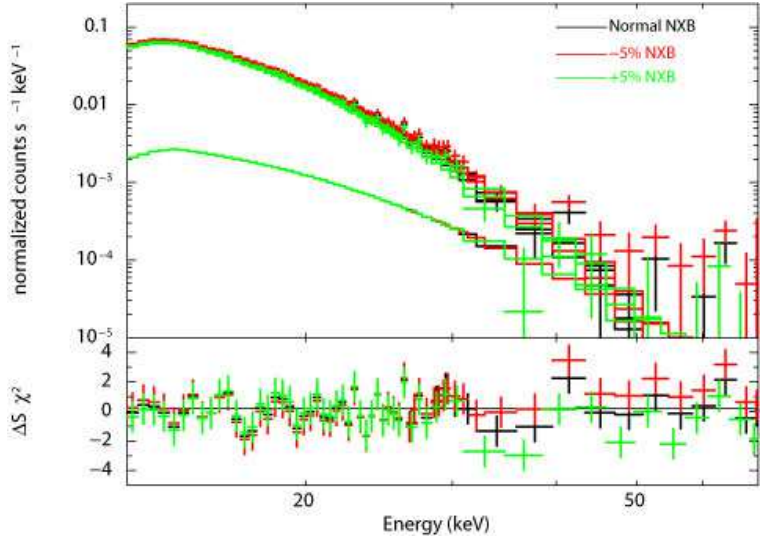}
  \end{center}
  \caption{HXD PIN spectra, including the effect of varying the
  normalization of the NXB by $\pm$5\% (crosses).  The NXB has been
  subtracted. The fit (solid line) is for the normal NXB. The lower
  panel plots the residuals divided by the $1\sigma$
  errors.}\label{fig:hxd}
\end{figure}

We thus combine the five spectra into one and perform a fit again. The
NXB spectra and the ARF are combined, correspondingly.  The result of
the fit is $kT=9.0_{-0.3}^{+0.3}$~keV (table~\ref{tab:spec} and
figure~\ref{fig:hxd}).  The spectrum does not require an extra hard
X-ray component. The flux of the thermal component in 20--80 keV band is
$6.8\times 10^{-11}\rm\: erg\: s^{-1}\: cm^{-2}$, which is about 
20\% higher than that expected from the BeppoSAX PDS observations
\citep{nev04}.

We note that the systematic uncertainty in the HXD NXB strongly affects
the results.  According to \citet{miz07}, the systematic $1\sigma$
uncertainty in the 15--40~keV NXB is 3.2~\%.  We conservatively adopt a
systematic uncertainty of $\pm$5\% (90\% uncertainty), and estimate the
upper limit on any extra hard X-ray component. Since the uncertainty in
the NXB overwhelms the statistical error on the ICM temperature, we fix
the temperature of the APEC model at 9.0~keV and add a power-law
component to represent the extra hard X-ray emission. We fixed the
photon index of the power-law at 2.0, which is the same as that in
\citet{eck08}. The 90\% upper limit on the hard power-law component is
obtained by employing the NXB spectra for which normalization is 5\%
lower than the normal NXB. The HXD spectra including $\pm$5\% variations
in the normalization of the NXB are shown in figure~\ref{fig:hxd}.  The
upper limit on the flux of any extra hard X-ray component is $2.8\times
10^{-11}\rm\: erg\: s^{-1}\: cm^{-2}$ (20--60~keV), which is consistent
with the detection obtained with INTEGRAL (0.7--$1.3\times 10^{-11}\rm\:
erg\: s^{-1}\: cm^{-2}$; \cite{eck08}) and the upper limit from a
combined Swift/BAT and Chandra spectrum ($5.2 \times 10^{-13}\rm\:erg\:
s^{-1}\: cm^{-2}$; \cite{aje08}).  If we assume that the temperature of
the APEC model is 8.5~keV as \citet{eck08} did, the upper limit on the
flux is $4.1\times 10^{-11}\rm\: erg\: s^{-1}\: cm^{-2}$
(20--60~keV). On the other hand, If we assume that the ICM temperature
is 9.7~keV (region~C; Table~\ref{tab:spec}), it is $1.0\times
10^{-11}\rm\: erg\: s^{-1}\: cm^{-2}$.  Note that the higher temperature
is closer to the value found by \citet{aje08}.

\section{Ionization Equilibrium}

% ## deleted
%If the Ophiuchus cluster is a dynamically young cluster, the ICM may not
%be in equilibrium. 
The high spectral resolution of Suzaku makes it possible to discuss the
ionization equilibrium state of the ICM by measuring the intensity ratio
of the hydrogen-like Fe K$\alpha$ line to the helium-like Fe K$\alpha$
line.  We limited the energy range to 5--8~keV. The spectral components
other than the ICM are the same as those in section~\ref{sec:spec}. We
fit the ICM component with a Bremsstrahlung model and two gaussians. The
spectra of the three XISs are fitted simultaneously.

\begin{figure}
  \begin{center}
    \FigureFile(100mm,100mm){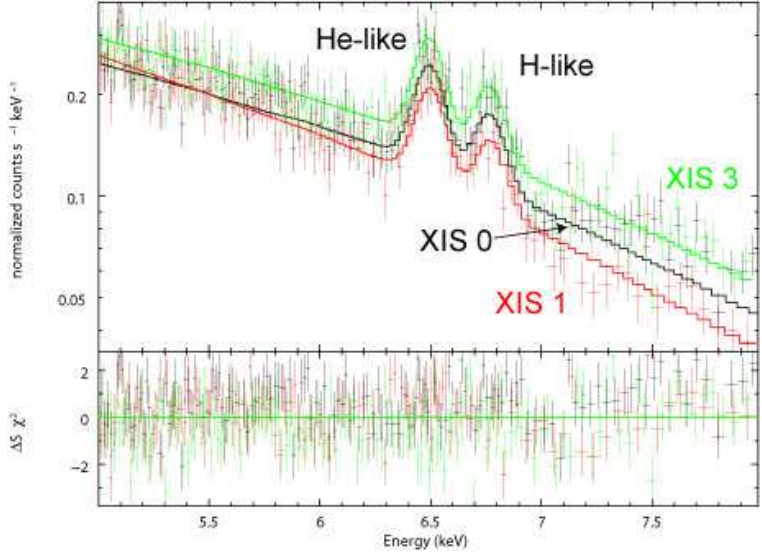}
  \end{center}
  \caption{XIS spectra around the Fe K$\alpha$ lines (crosses) and the
  results of the fit (solid lines) for region C-SE. The NXB spectra have
  been subtracted. The lower panel plots the residuals divided by the
  $1\sigma$ errors.}\label{fig:line}
\end{figure}

\begin{figure}
  \begin{center}
    \FigureFile(80mm,80mm){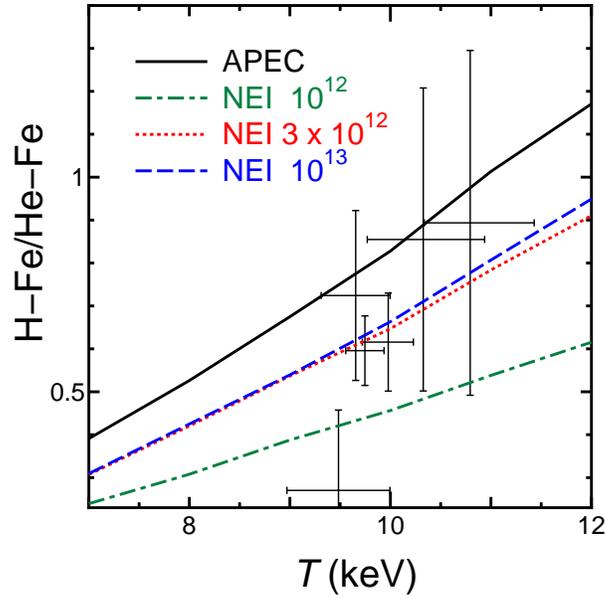}
  \end{center}
  \caption{Observed ratios of H-like and He-like Fe K$\alpha$ lines
 (crosses). The prediction of the APEC model is shown with the solid
 line, and those of the NEI model are shown with the dot-dashed line
 ($n_e t = 1\times 10^{12}\:\rm s\: cm^{-3}$), the dotted line ($n_e t =
 3\times 10^{12}\:\rm s\: cm^{-3}$) and the dashed line ($n_e t =
 1\times 10^{13}\:\rm s\: cm^{-3}$)}\label{fig:ratio}
\end{figure}

Figure~\ref{fig:line} shows the XIS spectra around the iron lines in
region C-SE. Figure~\ref{fig:ratio} shows the relation between the line
ratios and the temperatures listed in table~\ref{tab:spec}. The data
include those for the three central regions (C-NW, C-SW, C-SE) and
regions N, S, and E; the ratio for region W cannot be constrained, and
that for region C-NE is omitted because this region is multi-temperature
(figure~\ref{fig:depro}a). For comparison, we present the relation
predicted by the APEC model and those predicted by the NEI model (Ver.\
2), which is based on the Astrophysical Plasma Emission Database (APED;
\cite{smi01}). While APEC is a model for plasmas in ionization
equilibrium, NEI is a model for those not in equilibrium. In
figure~\ref{fig:ratio}, the observational data are generally consistent
with the APEC model.  The NEI model indicates $n_e t\gtrsim 3\times
10^{12}\:\rm s\: cm^{-3}$, where $n_e t$ is the ionization parameter.
The results of NEI as well as APEC indicate that the ICM has reached
ionization equilibrium.  They may also indicate that the ICM except for
the core consists of a single-temperature plasma and the cluster has no
region strongly locally heated by a merger (see \cite{ota08}). Although
$n_e t=1\times 10^{13}\:\rm s\: cm^{-3}$ is almost equivalent to
ionization equilibrium for NEI, there is a small difference between NEI
and APEC (figure~\ref{fig:ratio}). This may be due to the difference of
the atomic physics each model assumed.

\section{Discussion}

\subsection{Comparison with ASCA Results}

Although the Suzaku results are consistent with the previous ASCA
results for the central region of the cluster (table~\ref{tab:spec}),
they are not for the outer regions. The results of ASCA indicated that
the Ophiuchus cluster has a large temperature variation
\citep{wat01}. In particular, the temperatures of the western and
southern regions were reported to be extremely high ($kT\sim
15$--20~keV). On the other hand, our Suzaku results indicate that the
cluster is almost isothermal ($kT\sim 9$--10~keV).

One possible cause for the difference would be a problem with the X-ray
background for the ASCA observation.  While \citet{wat01} used a
background obtained from a source-free region near the cluster
(H~1705--250), we constructed a background based on a model of the
Galactic emission$^1$ plus a model for the hard CXB.  Since the western
and southern regions are dim, the difference of the background could
affect the results. However, we found that this is not likely to be the
case. We observed the region around H~1705--250 with Suzaku with an
exposure time of 15~ks. Then, we used the spectrum of this region as the
background for a spectral fit of region W. In this case, the temperature
is $kT=9.7_{-1.0}^{+0.9}$~keV, which is consistent with that in
table~\ref{tab:spec}. Another possibility is that the high temperature
in these faint outer regions is due to stray light from the bright
central regions of the cluster. For the X-ray telescopes carried on
ASCA, such problem was much more serious for ASCA than for Suzaku
\citep{ser07}. Although \citet{wat01} took account of the effect in
their analysis, the correction might not have been sufficient.

Although we could not find the cause of the inconsistency, we believe
that the results of Suzaku are more reliable, 
because the problem of the stray light has been much improved.

\subsection{The Cool Core}
\label{sec:cc}

The existence of a cool core (section~\ref{sec:depro}) and the
isothermality of the cluster (section~\ref{sec:spec}) indicate that the
Ophiuchus cluster is a cool core cluster. While many of clusters with
medium temperatures ($\sim 5$~keV) are cool core clusters having short
central cooling times, most of the hottest clusters with temperatures
similar to the Ophiuchus ($\gtrsim 10$~keV) do not seem to have cool
cores (e.g. figure~18c,d in \cite{ota04}). This is probably because the
latter are major-merger clusters in which the core has been destroyed
and the temperature has been boosted by the merger \citep{ran02}. For
example, the extended HIFLUGCS sample includes 106 nearby clusters
\citep{rei02}. Among them, there are only 8 clusters, including the
Ophiuchus cluster, with temperatures larger than 9.0~keV
\citep{che07}. Five of them (A~1689, A~1914, A~2163, A~2319, and
Triangulum Australis) do not have a prominent cool core
\citep{and04,gov04,whi00}. Although A~754 and A~2142 have a cool core,
they are now undergoing a major merger \citep{mar00,gov04}. Thus, the
Ophiuchus cluster seems to be exceptional in having a very high
temperature and a cool core with no strong evidence of a major
merger. Since clusters with cool cores tend to be old
(e.g. \cite{fuj00b,ota06}), the Ophiuchus cluster would be exceptionally
old given its large mass. Contrary to merging clusters in which star
formation activities might be induced in the disk galaxies (e.g.,
\cite{fuj99a,fuj99}), the environment in the Ophiuchus cluster might
have been calm for a long time. Thus, it would be interesting to study
the nature of the galaxies inside it.

\subsection{Non-Thermal Hard X-ray Emission}

Most of the clusters from which non-thermal X-ray emission has been
detected are major-merger clusters \citep{nev04}. Since non-thermal
synchrotron radio emission has often been observed from these clusters
\citep{gov04b}, the non-thermal hard X-ray emission could be attributed
to inverse Compton scattering of photons by the relativistic electrons
that are responsible for the radio emission. Since the cooling time of
these electrons ($\sim 10^8$~yr) is much shorter than the dynamical
time-scale of a cluster ($\sim 10^9$~yr; \cite{sar99}), the electrons
are likely to be primary electrons, which are now being directly
accelerated at shocks or turbulence in the ICM induced by recent cluster
mergers.

The hard X-ray excess detected by \citet{eck08} from the Ophiuchus
cluster is curious in this regard. Although a cluster merger does not
always destroy a cool core (e.g. A~2142; \cite{mar01}) and our
temperature map (figure~\ref{fig:TZ}a) is coarse, the X-ray image and
spectra from this cluster do not show any strong evidence for a major
merger. One possibility is that the Ophiuchus cluster is currently
undergoing an early stage, small impact parameter (head-on) merger with
the merger axis being close to our line of sight.  Then, the two merging
subclusters would be projected on top of one another, and no azimuthal
temperature variations would necessarily be expected.  By the same
argument, one would not necessarily expect any azimuthal variations in
the mean redshift. Although the likelihood of a head-on collision along
the line of sight is small, this hypothesis would appear to be
consistent with the apparent regularity of Ophiuchus, the presence of a
cool core (albeit without any very cool gas), and the presence of
nonthermal hard X-ray emission due to a merger shock or turbulence.

On the other hand, if Ophiuchus really is not a merging cluster, then
the nonthermal hard X-rays cannot be attributed to primary electrons
accelerated through a recent major cluster merger, although we cannot
rule out particle-acceleration by turbulence in the core generated by a
minor merger \citep{fuj04,fuj05,maz08}. One possibility is that
particles might have been accelerated by the central AGN rather than a
cluster merger, because the hard X-ray emission is observed at the
center of the Ophiuchus cluster \citep{eck08}. Moreover, since recent
strong AGN activities have not been observed at the cluster center
\citep{dun06}, the non-thermal emission from the cluster may come from
secondary electrons that are generated through proton-proton interaction
\citep{pfr04}. In this case, the non-thermal emission can be observed
even long after the AGN activities and proton acceleration cease,
because the cooling time of the protons is longer than the age of a
cluster. Recently, \citet{fuj07c} proposed that the protons are
accelerated at the forward shock around a cocoon created by an AGN
outburst.

Another possible source of the non-thermal X-ray emission would be dark
matter annihilation (\cite{pro08}, see also \cite{tot04}). In any case,
a confirmation of the non-thermal emission in Ophiuchus with other
instruments would also be very useful.  The Swift/BAT instrument
observed Ophiuchus for 1.3 Msec, and failed to detect any non-thermal
hard X-ray emission \citep{aje08,oka08}. Based on a joint Chandra and
Swift/BAT analysis (with Chandra used to constrain the thermal emission
from the cluster), \citet{aje08} found an upper limit of $5.2 \times
10^{-13}\rm\:erg\: s^{-1}\: cm^{-2}$ (20--60 keV), which is inconsistent
with the INTEGRAL detection at more than 2$\sigma$.  We note that
\citet{eck08} fit the thermal emission from the cluster with with a
single temperature model with a temperature of 8.5~keV.  We find that
most of the cluster is hotter than this, in agreement with the results
from \citet{aje08} and \citet{oka08}. The difference with \citet{eck08}
may be due to the cool core in Ophiuchus.  It is possible that a more
complex thermal model may be needed for Ophiuchus, and might bring the
INTEGRAL and Swift/BAT results into better agreement.  Given the present
disagreement of the INTEGRAL and Swift/BAT spectral results, the lack of
evidence for a merger, and the lack of a modern radio detection of a
halo or relic, we view the non-thermal detection in Ophiuchus as
tentative.

\section{Summary}

We observed the Ophiuchus cluster with Suzaku X-ray satellite. We found
that the ICM of the cluster is hot ($kT\sim 9$--10~keV) and is almost
isothermal except for the core. Contrary to a previous study, we did not
find extremely hot regions ($kT\sim 20$~keV) in the cluster. In the
core, the temperature is smaller and the metal abundance is larger than
those in the surrounding region. The radial velocity difference between
different regions of the ICM is less than $\sim 3000\rm\: km\: s^{-1}$.
The iron line ratios in X-ray spectra suggest that the ICM has reached a
ionization equilibrium. These features indicate that the Ophiuchus
cluster is not a major-merger cluster but is a so-called cool core
cluster. This cluster would be an exceptional cluster because most of
the clusters with these temperatures ($\gtrsim 10$~keV) are major-merger
clusters. Another possibility is that Ophiuchus is a nearly head-on
merger with the merger axis nearly along our line of sight; such a model
would be consistent with the observed properties of the cluster. We
obtained the upper limit of hard X-ray excess, which is consistent with
the recent detection with INTEGRAL and upper limit with Swift/BAT.
Assuming that the Ophiuchus cluster is not a major-merger cluster, the
origin of any hard X-ray excess may not be a recent major cluster
merger.

\vspace{5mm}

The authors wish to thank A. Furuzawa, R. K. Smith, and N. Ota for
useful comments.  We also thank the Suzaku operations team for their
support in planning and executing these observations.  Y.~F., K.~H., and
M.~T. were supported in part by a Grant-in-Aid from the Ministry of
Education, Culture, Sports, Science, and Technology of Japan (Y.~F.:
17740162; K.~H.: 16002004; M.~T.: 16740105,
19740096). T.~H.~R. acknowledges support by the Deutsche
Forschungsgemeinschaft through Emmy Noether Research Grant RE 1462.
C.~L.~S. was supported by NASA Suzaku grants NNX06AI37G and NNX06AI44G.

%%%
% See the manual for the detail.
%%%

\end{document}